\documentclass[aps,prb,showpacs,twocolumn]{revtex4-1}
\usepackage{amssymb}
\usepackage{amsmath}
\usepackage{graphicx}
\usepackage{dsfont}
\usepackage{times}

\begin{document}

\title{Random Matrix Model for Eigenvalue Statistics in Random Spin Systems}
\author{Wen-Jia Rao}
\email{wjrao@hdu.edu.cn}
\affiliation{School of Science, Hangzhou Dianzi University, Hangzhou 310027, China.}
\date{\today }

\begin{abstract}
We propose a working strategy to describe the eigenvalue statistics of
random spin systems along the whole phase diagram with thermal to many-body
localization (MBL) transition. Our strategy relies on two random matrix (RM)
models with well-defined matrix construction, namely the mixed (Brownian)
ensemble and Gaussian $\beta$ ensemble. We show both RM models are capable
of capturing the lowest-order level correlations during the transition,
while the deviations become non-negligible when fitting higher-order ones.
Specifically, the mixed ensemble will underestimate the longer-range level
correlations, while the opposite is true for $\beta$ ensemble. Strikingly, a
simple average of these two models gives nearly perfect description of the
eigenvalue statistics at all disorder strengths, even around the critical
region, which indicates the interaction range and strength between
eigenvalue levels are the two dominant features that are responsible for the
phase transition.
\end{abstract}

\maketitle

\section{Introduction}

\label{sec1}

Many-body localized (MBL) phase\cite{Gornyi2005,Basko2006}, as the only
example of phase that violates eigenstate thermalization hypothesis (ETH)%
\cite{Deutsch,Srednicki} in an isolated quantum system, is a focus of
current condensed matter physics. Modern understanding about MBL phase and
its counterpart -- thermal phase that respects ETH -- relies on quantum
entanglement. Specifically, thermal phase is ergodic with delocalized
eigenstate wavefunctions, which results in extensive (volume law)
entanglement between subsystems. On the contrary, MBL phase is signatured by
small (area-law) entanglement. The qualitative difference in the scaling of
quantum entanglement is widely used in the study of thermal-MBL transition
\cite%
{Kjall2014,Geraedts2017,Yang2015,Serbyn16,Gray2017,Maksym2015,Kim,Bardarson,Abanin}
.

More traditionally, the thermal and MBL phase are distinguished by their
eigenvalue statistics\cite%
{Oganesyan,Avishai2002,Regnault16,Regnault162,Huse1,Huse2,Huse3,Garcia,Luitz},
whose mathematical foundation is laid by the random matrix (RM) theory\cite%
{Mehta,Haake2001}. The eigenvalues of thermal phase are well-correlated,
whose statistics divides into three Wigner-Dyson (WD) classes depending on
the system's symmetry: the Gaussian orthogonal ensemble (GOE) for orthogonal
systems with time reversal symmetry, the Gaussian unitary ensemble (GUE) for
those break time reversal symmetry, and Gaussian symplectic ensemble (GSE)
for time-reversal invariant systems with broken spin rotational invariance.
On the contrary, the eigenvalues in MBL phase are independent of each other
and follows Poisson statistics. Quantitatively, the eigenvalue statistics is
evaluated by the distribution of ratios between two adjacent level spacings
(gaps)%
\begin{equation}
r_{i}^{\left( n\right) }=\frac{E_{i+2n}-E_{i+n}}{E_{i+n}-E_{i}}\text{.}
\end{equation}%
The lowest-order one $n=1$ was proposed in Ref.[\onlinecite{Oganesyan}] as
the standard probe for nearest level correlation, and higher order ones with
$n>1$ describe level correlations on longer ranges. Compared to the more
traditional quantity like level spacing $\left\{ s_{i}^{\left( n\right)
}=E_{i+n}-E_{i}\right\} $ or number variance $\Sigma ^{2}$, spacing ratios
are independent of density of states and requires no unfolding procedure,
which is non-unique and may raise subtle misleading signatures in certain
models\cite{Gomez2002}.

Besides the level statistics deep in the thermal/MBL phase, there are also
significant amount of works on the spectral statistics right at the critical
point, or even along the whole phase diagram\cite%
{Shukla,Serbyn,SRPM,Mix,Sierant19,Rao21,Sierant20,Buijsman}. For example,
the single-parameter Gaussian $\beta $ ensemble, which generalizes the
standard Gaussian ensembles into the one with continuous Dyson index.
However, as we shall see, it can not accurately account for the high-order
level statistics, especially for the MBL phase. As a generalization, the
two-parameter $\beta -h$ model, recently proposed in Ref.[%
\onlinecite{Sierant20}], was shown to reproduce $P\left( r^{\left( n\right)
}\right) $ with high accuracy during the MBL transition, which indicates the
interaction strength and range are the two dominant varying features along
with the phase transition. However, the $\beta -h$ model is based on the
joint probability distribution of eigenvalues, and the two parameters $\beta
$ and $h$ has to be determined jointly, which is numerically difficult to
achieve. This motivates us to search for a RM model that based directly on
the matrix construction to reproduce the level statistics -- both on short
and long ranges -- of random spin systems.

In this work, we propose another working strategy to reproduce $P\left(
r^{\left( n\right) }\right) $ along the thermal-MBL transition in 1D random
spin systems. Our strategy is based on two RM models that have well-defined
parent matrix construction, i.e. the mixed ensemble (also called Brownian
ensemble\cite{Shukla2000,Shukla2005} or Rosenzweig-Porter ensemble\cite%
{Mehta,Shapiro,Kravtsov2015} in the literature) and Gaussian $\beta $
ensemble, both of which incorporate the WD and Poisson distribution in a
direct manner. We will show both RM models can accurately reproduce $P\left(
r^{\left( 1\right) }\right) $ with properly chosen model parameters, but
they both show non-negligible deviations when fitting higher-order spacing
ratios. Specifically, the mixed ensemble will \emph{underestimate} the
longer-range level correlations, while the opposite is true for $\beta $
ensemble. Surprisingly, an average distribution of these two models gives
nearly perfect description for the level statistics on moderate ranges along
the whole phase diagram, even at the critical region. We argue these results
also suggest the strength and range of level interaction are responsible for
the MBL transition, in agreement with the conclusions of earlier works\cite%
{Corps,Sierant20}.

This paper is organised as follows. Sec.\ref{sec2} introduces the mixed
ensemble and Gaussian $\beta $ ensemble. In Sec.\ref{sec3} we use these
models to fit $P\left( r^{\left( n\right) }\right) $ in an orthogonal random
spin system. Particularly, we will introduce the average distribution of the
two RM models, which is shown to give fairly good descriptions of $P\left(
r^{\left( n\right) }\right) $ with $n>1$, even at the transition region. In
Sec.\ref{sec4} we verify this strategy in random spin systems with unitary
symmetry and quasi-periodic potential. Conclusion and discussion come in
Sec. \ref{sec5}.

\section{Random Matrix Models}

\label{sec2}

The basic requirement for an effective RM model for thermal-MBL transition
is that it should incorporate both WD (for thermal phase) and Poisson
statistics (for MBL phase). To this end, the first RM model we consider is
the mixed ensemble

\begin{equation}
M_{\alpha \rightarrow 0}\left( x\right) =xM_{\alpha }+\left( 1-x\right)
M_{0}.  \label{equ:ME}
\end{equation}%
where $M_{\alpha }$ with Dyson index $\alpha =1,2,4$ represents matrix in
the WD class, $M_{0}$ is a diagonal matrix with random diagonals standing
for Poisson ensemble, and the normalization condition is chosen to be $%
Tr\left( M_{\alpha /0}^{2}\right) =1$. It's easy to see the eigenvalue
statistics of $M_{\alpha \rightarrow 0}\left( x\right) $ evolves from
Poisson to WD when $x$ ranges from $0$ to $1$. Unfortunately, $P\left(
r^{\left( n\right) }\right) $ in the mixed ensemble lacks a compact
analytical expression\cite{Schierenberg,Chavda,Corps2}, so we will use
numerical results instead. Specifically, we numerically generate samples of
eigenvalue spectrum of Eq. (\ref{equ:ME}) in the range $x\in \left(
0,1\right) $ with interval $dx=0.01$, where the matrix dimension and sample
number are kept to be $1000$. After sampling, we take $400$ eigenvalues in
the middle of each spectrum to determine $P\left( r^{\left( n\right)
}\right) $, which will be used for future fittings.

The second considered RM model is the Gaussian $\beta $ ensemble, which is
an generalization of the standard WD ensembles into the one with a
continuous Dyson index $\beta \in \left( 0,\infty \right) $, whose joint
probability distribution of the eigenvalues is,%
\begin{equation}
P\left( \left\{ E_{i}\right\} \right) \propto \prod_{i<j}\left\vert
E_{i}-E_{j}\right\vert ^{\beta }e^{-\beta \sum_{i}E_{i}^{2}/2}\text{.}
\end{equation}%
The generalized Dyson index $\beta $ essentially controls the strength of
level repulsion, and the limit $\beta \rightarrow 0$ stands for the Poisson
ensemble with uncorrelated eigenvalues. The $\beta $ ensemble can be
generated by a tridiagonal RM\cite{Beta}%
\begin{equation}
M_{\beta }=\frac{1}{\sqrt{2}}\left(
\begin{array}{ccccc}
x_{1} & y_{1} &  &  &  \\
y_{1} & x_{2} & y_{2} &  &  \\
&
\begin{array}{ccc}
\text{.} &  &  \\
& \text{.} &  \\
&  & \text{.}%
\end{array}
&
\begin{array}{ccc}
\text{.} &  &  \\
& \text{.} &  \\
&  & \text{.}%
\end{array}
&
\begin{array}{ccc}
\text{.} &  &  \\
& \text{.} &  \\
&  & \text{.}%
\end{array}
&  \\
&  & y_{N-2} & x_{N-1} & y_{N-1} \\
&  &  & y_{N-1} & x_{N}%
\end{array}%
\right)  \label{equ:Beta}
\end{equation}%
where the diagonals $x_{i}\,$\ follow the normal distribution $\mathit{N}%
\left( 0,2\right) $ and $y_{k}$ ($k=1,2,...,N-1$) follows the $\chi $
distribution with parameter $\left( N-k\right) \beta $. For the $\beta $
ensemble, analytical and strong numerical evidences support $P\left(
r^{\left( n\right) }\right) $ to have the following compact form\cite%
{Atas,Tekur,Rao20,Rao202}
\begin{eqnarray}
P\left( \beta ,r^{\left( n\right) }=r\right) &=&Z_{\beta }\frac{\left(
r+r^{2}\right) ^{\gamma }}{\left( 1+r+r^{2}\right) ^{1+3\gamma /2}}\text{,}
\label{equ:1} \\
\gamma &=&\frac{n\left( n+1\right) }{2}\beta +n-1
\end{eqnarray}%
where $Z_{\beta }$ is the normalization factor determined by $%
\int_{0}^{\infty }P\left( \beta ,r^{\left( n\right) }\right) dr^{\left(
n\right) }=1$. This model has been used to describe the level statistics in
random spin systems\ in Ref.[\onlinecite{Buijsman}], where the authors
showed that $\beta $ ensemble is capable of capturing lowest-order spacing
ratio distributions during thermal-MBL transition, while the fittings for
higher-order ones have non-negligible deviations. In this work, we will not
only confirm this conclusion, but also show how these deviations can be
fixed.

\bigskip Both the mixed ensemble and $\beta $ ensemble incorporate the
transition from WD to Poisson with tuning parameters $x$ and $\beta $, but
they have sharp difference, which is easiest to see from the viewpoint of
level dynamic. By mapping the eigenvalues of RM into an one-dimensional
system of interacting classical particles, the joint probability
distribution of the former can be written into the canonical ensemble
distribution of the latter, that is,%
\begin{eqnarray}
P\left( \left\{ E_{i}\right\} ,\beta \right) &=&Z_{\beta }^{-1}e^{-\beta
H\left( \left\{ E_{i}\right\} \right) }\text{,} \\
H\left( \left\{ E_{i}\right\} \right) &=&\sum_{i}U\left( E_{i}\right)
+\sum_{\left\vert i-j\right\vert <h}V\left( \left\vert
E_{i}-E_{j}\right\vert \right) \text{,}
\end{eqnarray}%
where $U\left( E_{i}\right) \propto E_{i}^{2}$ is the background trapping
potential, and $V\left( \left\vert E_{i}-E_{j}\right\vert \right) $ controls
the level correlations. It's easy to see the choice with $V\left( x\right)
\propto \log \left\vert x\right\vert $ and interaction range $h\rightarrow
\infty $ corresponds to the Gaussian $\beta $ ensemble, where the Dyson
index is interpreted as the inverse temperature (or equivalently, the
interaction strength).

By this mapping, there are three aspects that determine the level
statistics: the form of level interaction $V\left( x\right) $, the
interaction range $h$ and strength $\beta $. For the mixed ensemble, all the
three aspects will change when varying $x$; while in the Gaussian $\beta $
ensemble, only the interaction strength $\beta $ can change. It is then the
key question that whether all the three aspects contribute in a physical
thermal-MBL transition, or whether only one or two of them do. We will
explore this question in random spin systems.

\section{Orthogonal Spin Chain}

\label{sec3}

We consider the canonical system for MBL, that is, the one-dimensional spin-$%
1/2$ chain with random external fields, whose Hamiltonian is\cite{Alet}%
\begin{equation}
H=\sum_{i=1}^{L}\mathbf{s}_{i}\cdot \mathbf{s}_{i+1}+\sum_{\alpha
=x,y,z}h_{\alpha }\sum_{i=1}^{L}\varepsilon _{i}^{\alpha }s_{i}^{\alpha }
\label{equ:Ham}
\end{equation}%
where periodic bounrady condition is imposed in the Heisenberg term, and $%
\varepsilon _{i}^{\alpha }$s are random numbers in $\left[ -1,1\right] $. We
consider here the orthogonal case that $h_{x}=h_{z}=h$ and $h_{y}=0$, which
is known to exhibit a thermal-MBL transition at around $h_{c}\simeq 3$\cite%
{Regnault16,Regnault162}, with the corresponding level statistics evolving
from GOE to Poisson. Compared to the more widely-studied case with $%
h_{x}=h_{y}=0$, our choice breaks total $S^{z}$ conservation and makes the
eigenstates fully featureless, hence is less affected by finite-size effects.

\begin{figure*}[t]
\centering
\includegraphics[width=2\columnwidth]{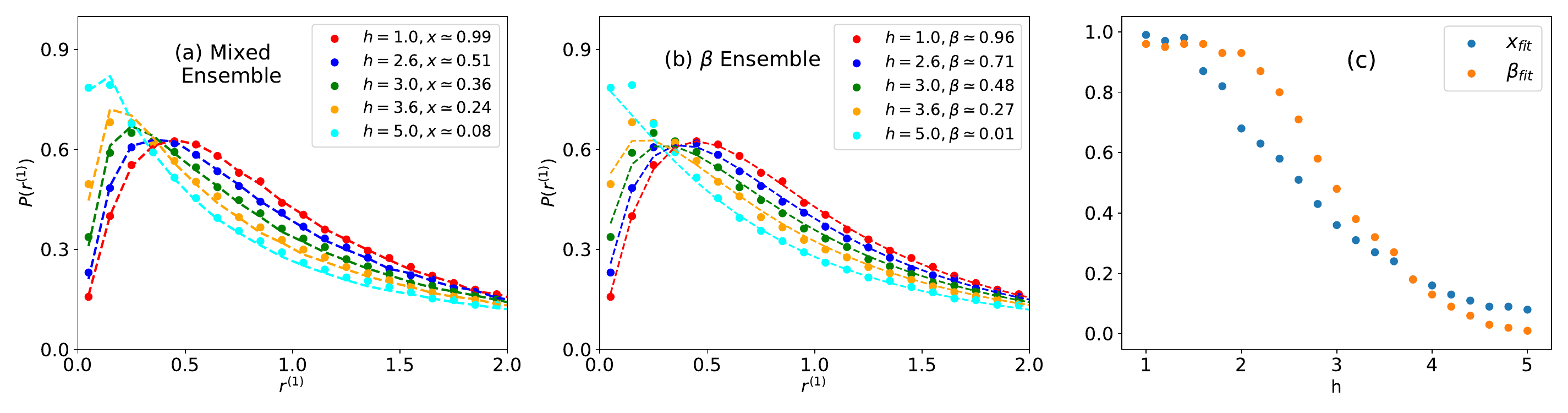}
\caption{The fittings of $P\left(r^{(1)}\right)$ in orthogonal spin chain by
(a) the mixed ensemble and (b) Gaussian $\protect\beta $ ensemble, where the
colored dots stand for numerical data of the physical Hamiltonian, and
dotted lines stand for the optimized fitting results from RM models, the
fitted model parameters are displayed in the figure legends. (c) Evolutions
of the fitted parameters $x_{\text{fit}}$ and $\protect\beta_{\text{fit}}$
along the thermal-MBL transition.}
\label{fig:OrthNN}
\end{figure*}

For the first step, we will show both the mixed ensemble and Gaussian $\beta
$ ensemble can accurately reproduce $P\left( r^{\left( 1\right) }\right) $
with properly chosen model parameters. These proper parameters are found by
minimizing the following distance%
\begin{equation}
D\left( \lambda \right) =\sum_{r}\left\vert P_{\text{mol}}\left( \lambda
,r^{\left( 1\right) }\right) -P_{\text{num}}\left( r^{\left( 1\right)
}\right) \right\vert ^{2}\text{,}  \label{fig:Dist}
\end{equation}%
where $P_{\text{mol}}\left( \lambda ,r^{\left( 1\right) }\right) $ is the
target model distribution with parameter $\lambda $ standing for $x$ ($\beta
$) in the mixed ensemble ($\beta $ ensemble), and $P_{\text{num}}\left(
r^{\left( 1\right) }\right) $ is the numerical data from physical model. For
the latter, we simulate Eq. (\ref{equ:Ham}) in an $L=13$ system, with the
Hilbert space dimension $N_{d}=2^{13}=8192$, and generate $400$ samples of
eigenvalue spectrum at each disorder strengths. For each sample, we select $%
400$ eigenvalues in the middle to determine $P_{\text{num}}\left( r^{\left(
1\right) }\right) $. We then determine the proper parameters by minimizing $%
D\left( \lambda \right) $, and draw the resulting $P_{\text{mol}}\left(
r^{\left( 1\right) }\right) $ together with $P_{\text{num}}\left( r^{\left(
1\right) }\right) $, the results are collected in Fig.~\ref{fig:OrthNN}%
(a),(b).

As can be seen, both RM models reproduce $P_{\text{num}}\left( r^{\left(
1\right) }\right) $ to a satisfying accuracy at all disorder strengths, even
at the transition region. Generally, the mixed ensemble gives better
performance than the $\beta $ ensemble, especially for cases with large
disorder. This is because, when approaching the MBL phase, eigenvalue
correlations become significantly short-ranged, while $\beta $ ensemble
preserves level correlations on \emph{all} ranges, which gives rise to
larger deviations. These deviations will be more transparent when fitting
higher-order spacing ratios.

To be complete, we draw the evolution of proper model parameters $x_{\text{%
fit}}$ and $\beta _{\text{fit}}$ with respect to the randomness strength $%
h\in \left[ 1,5\right] $ in Fig.~\ref{fig:OrthNN}(c). We observe monotonic
decreasing tendencies for $x_{\text{fit}}$ and $\beta _{\text{fit}}$, both
of which stand for decreasing level correlations, in consistent with
physical intuition.

\begin{figure}[t]
\centering
\includegraphics[width=\columnwidth]{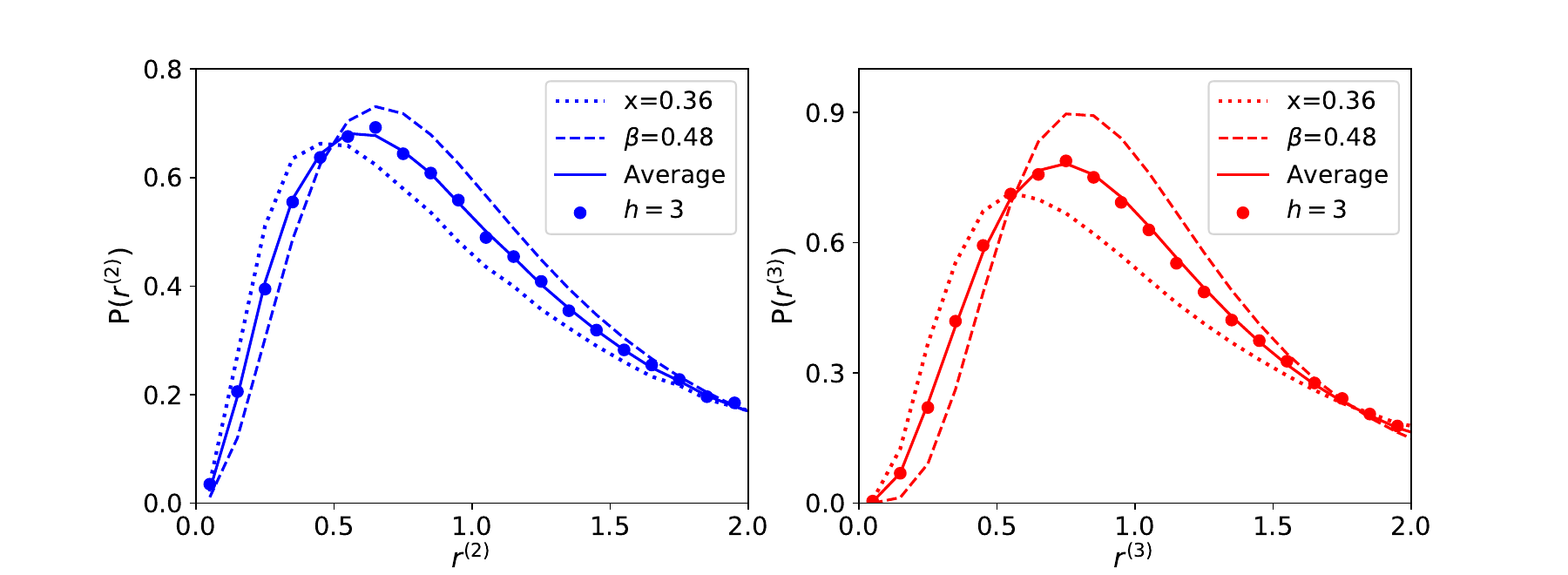}
\caption{The fittings for the (left) 2nd-order spacing ratio $P\left( r^{\left(
2\right) }\right)$ and (right) 3rd-order spacing ratio $P\left( r^{\left(
3\right) }\right) $ for physical model with randomness strength $h=3$. The dotted and dashed lines correspond to the mixed ensemble
and $\beta$ ensemble respectively, whose model parameters are determined
by fitting $P\left( r^{\left( 1\right)}\right) $, the solid lines are drawn by averaging the dotted and dashed lines.
} \label{fig:NNN}
\end{figure}

Now we proceed to study the longer-range level correlations through the
higher-order spacing ratios $P\left( r^{\left( n\right) }\right) $ with $n>1$%
. To get an intuitive picture, we take the case with $h=3$ as a
demonstration, which is at the critical region with largest fluctuations.
The proper parameters can be read from Fig.~\ref{fig:OrthNN}(c), which is $x_{%
\text{fit}}=0.36$ for the mixed ensemble and $\beta _{\text{fit}}=0.48$ for
the $\beta $ ensemble. We then draw the corresponding $P\left( r^{\left(
2\right) }\right) $ and $P\left( r^{\left( 3\right) }\right) $ of both
models, and compare them to the physical data, the results are shown in Fig.~\ref{fig:NNN}%
. As we can see, both RM models show non-negligible deviations. More
specifically, the peak of $P\left( r^{\left( 2/3\right) }\right) $ in the $%
\beta $ ensemble is higher than the physical data, meaning it \emph{%
overestimates} the longer-range level correlations, while the opposite is
true for the mixed ensemble.

Surprisingly, if we take a closer look at Fig.~\ref{fig:NNN}, we see the
physical data lies roughly at the middle of the mixed ensemble and $\beta $
ensemble, which motivates us to draw the average of them, that is%
\begin{equation}
P_{\text{ave}}\left( r^{\left( n\right) },x_{\text{fit}},\beta _{\text{fit}%
}\right) =\frac{P_{\text{mix}}\left( x_{\text{fit}}\text{,}r^{\left(
n\right) }\right) +P_{\beta }\left( \beta _{\text{fit}}\text{,}r^{\left(
n\right) }\right) }{2}\text{.}  \label{equ:Ave}
\end{equation}%
The $P_{\text{ave}}\left( r^{\left( n\right) }\right) $ appear as the solid
colored lines in Fig.~\ref{fig:NNN}, they match almost perfectly with the
physical data, which indicates the deviations of two RM models cancel with
each other. To confirm this is not a coincidence, we draw $P_{\text{ave}%
}\left( r^{\left( n\right) }\right) $ up to $n=7$ at various disorder
strengths with corresponding proper parameters in Fig.~\ref{fig:NNN}(c), and
compare them to the physical data, the results are collected in Fig.~\ref%
{fig:OrthHigh}.

\begin{figure*}[t]
\centering
\includegraphics[width=1.9\columnwidth]{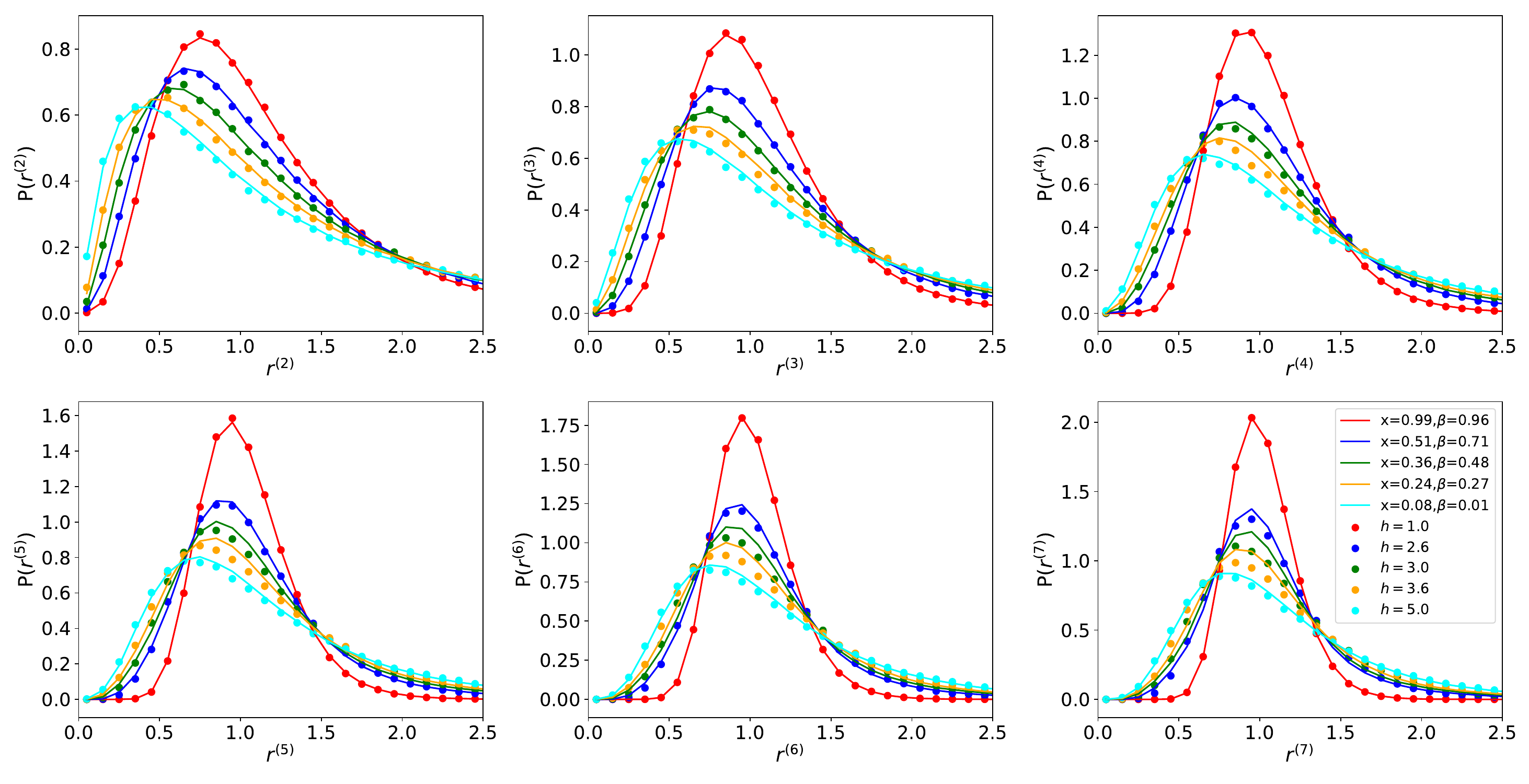}
\caption{The fittings for higher-order spacing ratios at various disorder
strengths, where the dots are numerical data, and lines are the average
distribution $P_{\text{ave}}\left( r^{\left( n\right) }\right) $ according
to Eq. (\protect\ref{equ:Ave}), all figures share the same figure legends.
Fairy good matches are found for $n\leq 4$, and the deviations starts to
grow for $n\geq 5$ in the critical region($h\simeq 3$), reflecting the
critical fluctuation in a finite system.}
\label{fig:OrthHigh}
\end{figure*}

As can be seen, $P_{\text{ave}}\left( r^{\left( n\right) }\right) $ meets
perfectly with the physical data up to $n=4$, that is, when $9$ consecutive
levels are concerned, even at the critical region ($h\simeq 3$). The
deviations starts to grow for $n>4$ at the critical region, which reflects
the large critical fluctuations. These results suggest $P_{\text{ave}}\left(
r^{\left( n\right) }\right) $ indeed gives an accurate account of the level
evolutions along the thermal-MBL transition, which works not only for
lowest-order level correlations, but also for correlations on moderate
longer ranges.

\bigskip Current results provide valuable indications about the evolution of
level dynamics during MBL transition. First, results in Fig.~\ref{fig:NNN}
indicate the $\beta $ ensemble overestimates the long-range level
correlations even it accurately captures the lowest-order one as shown
in Fig.~\ref{fig:OrthNN}(b). This is not surprising since $\beta $ ensemble
preserves level correlations on \emph{all} ranges, while the level
correlation becomes significantly short ranged when increasing disorder
strength. Actually, such deviations have already appeared when fitting $P\left(
r^{\left( 1\right) }\right) $ deep in MBL phase in Fig.~\ref{fig:OrthNN}(b),
which is further amplified when considering higher-order $P\left( r^{\left(
n\right) }\right) $. This indicates only the interaction strength $\beta $
is not sufficient to cover the eigenvalue evolution during MBL transition,
we must take interaction range into consideration. On the other hand, when
studying the mixed ensemble, the interaction form, interaction range and
strength all change when varying model parameter $x$, and results in a
underestimation of long-range level correlations. This fact indicates \emph{%
not} all the three aspects are responsible for MBL transition. Finally, an
average of the two RM models gives proper description of level correlations
along the MBL transition on moderate long ranges, which means the deviations
in individual RM model cancel with the other one. Therefore, the only
possible explanation is that the interaction between eigenvalues stays
logarithmic while the interaction strength and range change during MBL
transition, which in consistent with the $\beta -h$ model studied in Ref.[%
\onlinecite{Sierant20}]. However, as mentioned in the Introduction section,
the $\beta -h$ model is built on eigenvalue distributions, while our model
stems from two RM models with well-defined parent matrix construction.

The numerical results above are from an $L=13$ system, we have also
confirmed that $P_{\text{ave}}\left( r^{\left( n\right) }\right) $ works
fine for an $L=12$ system, although the fitted parameters $x_{\text{fit}}$
and $\beta _{\text{fit}}$ may have minor deviations, especially for cases in
the transition region. We suspect these fitted parameters should converge
when larger systems and more samples are considered, while the results from $%
L=13$ are sufficient to verify the efficiency of $P_{\text{ave}}\left(
r^{\left( n\right) }\right) $.

At this stage, a working three-step strategy to reproduce the eigenvalue
statistics at any disorder strength $h$ during thermal-MBL transition is
proposed as follows. Step 1: Numerically compute the lowest-order spacing
ratio distribution $P\left( r^{\left( 1\right) }\right) $ of the physical
Hamiltonian; Step 2: Find the proper parameter $x_{\text{fit}}$ (for the
mixed ensemble) and $\beta _{\text{fit}}$ (for the $\beta $ ensemble) that
fits best with $P\left( r^{\left( 1\right) }\right) $, which is done by
minimizing the distance $D\left(\lambda\right)$ in Eq. (\ref{fig:Dist});
Step 3: Compute the average distribution $P_{\text{ave}}\left( r^{\left(
n\right) }\right) $ according to Eq. (\ref{equ:Ave}) with the proper
parameters $x_{\text{fit}}$ and $\beta _{\text{fit}}$. The $P_{\text{ave}%
}\left( r^{\left( n\right) }\right) $ obtained in this manner is expected to
faithfully describe level statistics along the thermal-MBL transition, even
at critical region, at least when level correlations on moderate ranges are
concerned. To further support this strategy, we proceed to consider MBL
systems with unitary symmetry and quasi-periodic potential.

\section{Unitary and Quasi-Periodic Systems}

\label{sec4}

\bigskip To consider a unitary system, we simulate Eq. (\ref{equ:Ham}) with $%
h_{x}=h_{y}=h_{z}=h$ in an $L=13$ system, with the rest technical settings
identical to orthogonal case in previous section. This unitary model is
known to exhibit a thermal-MBL transition at $h_{c}\simeq 2.5$, with
corresponding RM description evolving from GUE to Poisson\cite%
{Regnault16,Regnault162}.

\begin{figure*}[t]
\centering
\includegraphics[width=2\columnwidth]{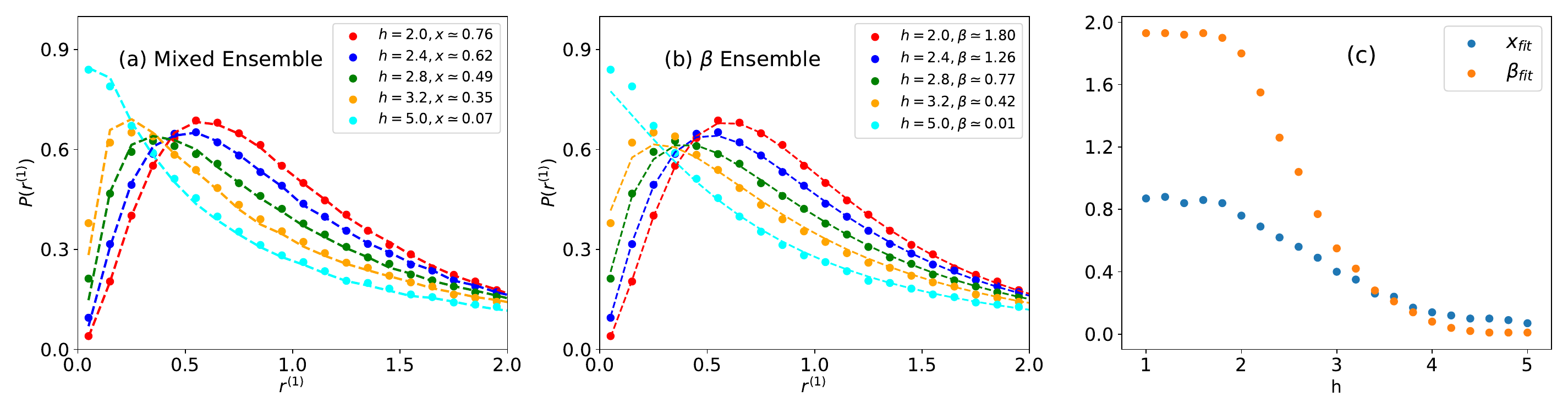}
\caption{The fittings of $P\left( r^{\left( 1\right) }\right) $ in the
unitary spin model by (a) the mixed ensemble and (b) Gaussian $\protect\beta
$ ensemble. (c) Evolution of the fitted parameter $x_{\text{fit}}$ and $%
\protect\beta _{\text{fit}}$ along the thermal-MBL transition.}
\label{fig:UnitNN}
\end{figure*}

\begin{figure*}[t]
\centering
\includegraphics[width=2\columnwidth]{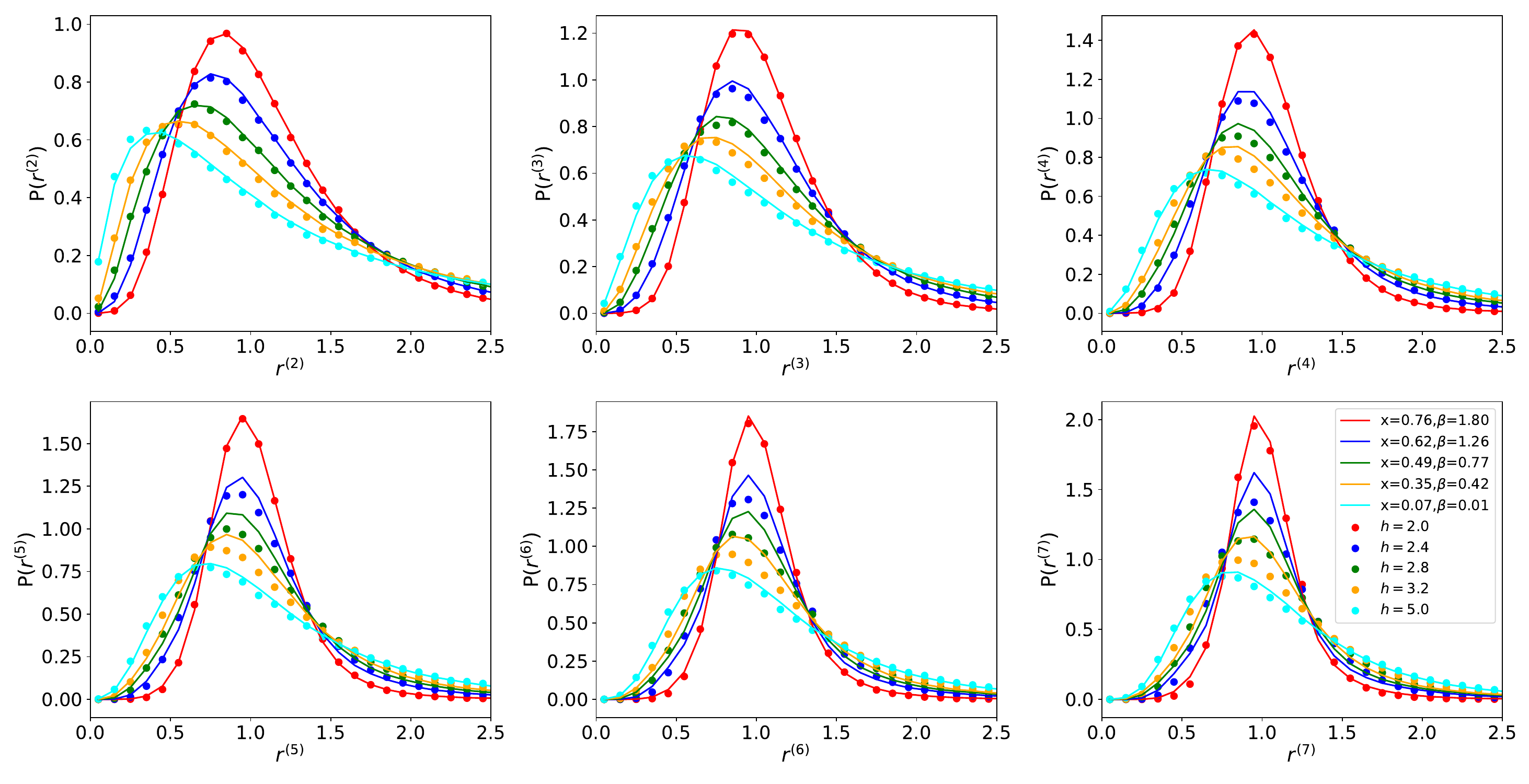}
\caption{The fittings of higher-order spacing ratios at various disorder
strengths in the unitary system, where the lines are the average
distribution $P_{\text{ave}}\left( r^{\left( n\right) }\right) $. The
fittings are close to perfect for $n\leq 3$, and starts to deviate for $%
n\ge4 $ for data in the critical region ($h\simeq2.5$). The deviations are
slightly lager than those in orthogonal system, reflecting larger critical
fluctuations in finite system.}
\label{fig:UnitHigh}
\end{figure*}

Like in the orthogonal case, we first show both the mixed ensemble and $%
\beta $ ensemble are capable of reproducing $P\left( r^{\left( 1\right)
}\right) $ with proper parameters $x_{\text{fit}}$ and $\beta _{\text{fit}}$
obtained by minimizing $D\left( \lambda \right) $ in Eq. (\ref{fig:Dist}).
The fitting results are in Fig.~\ref{fig:UnitNN}(a),(b), note the parameter $%
\beta $ is ranging from $2$ (GUE) to $0$ (Poisson), and $x$ is now a
parameter tuning the weight between GUE and Poisson. As expected, both RM
models reproduce $P\left( r^{\left( 1\right) }\right) $ quite well, even at
the critical region. The evolution of $x_{\text{fit}}$ and $\beta _{\text{fit%
}}$ are drawn in Fig.~\ref{fig:UnitNN}(c), where expected decreasing tendencies
are observed.

With the properly fitted parameters $x_{\text{fit}}$ and $\beta _{\text{fit}%
} $, we can determine $P_{\text{ave}}\left( r^{\left( n\right) }\right) $
according to Eq. (\ref{equ:Ave}) and compare them to the physical data at
various disorder strengths, the results are collected in Fig.~\ref%
{fig:UnitHigh}. As we can see, $P_{\text{ave}}\left( r^{\left( n\right)
}\right) $ meets perfectly with physical data at all disorder strengths up
to $n=3$, that is, when $7$ consecutive levels are considered. The
deviations starts to grow for $n\geq 4$ in the transition region. Compared
to the results in orthogonal model (perfect fittings for $n\leq 4$), the
deviations are slightly larger, which reflects the critical fluctuations are
larger in a unitary system.

\begin{figure*}[t]
\centering
\includegraphics[width=2\columnwidth]{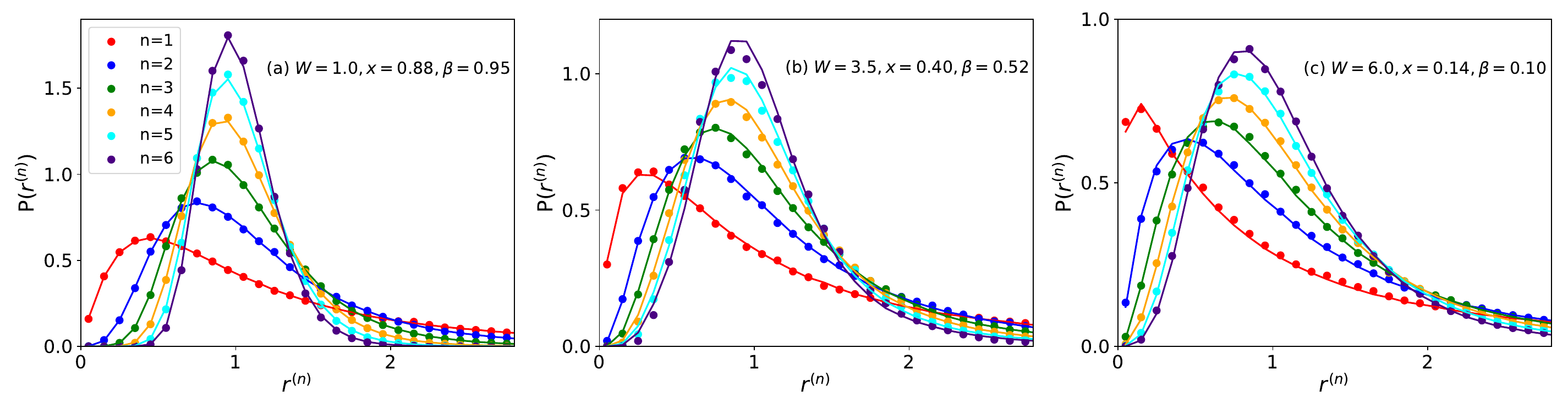}
\caption{The comparison between $P_{\text{ave}}\left( r^{\left( n\right)
}\right) $ got from Eq. (\protect\ref{equ:Ave}) and physical data in the QP
spin system in (a) thermal (W=1), (b) MBL (W=6) and (c) intermediate (W=3.5)
region, where the optimized parameters $x$ and $\protect\beta$ are obtained
by fitting $P\left( r^{\left( 1\right) }\right) $ with individual RM model.
The fittings are even better than those in random disorder systems, whose
qualitative explanations are in the main text.}
\label{fig:QP}
\end{figure*}

Furthermore, we test this strategy in an MBL system induced by a different
mechanism, that is, by quasi-periodic (QP) potential. The Hamiltonian is as
follow%
\begin{eqnarray}
H_{\text{QP}} &=&J\sum_{i=1}^{L-1}\mathbf{s}_{i}\cdot \mathbf{s}%
_{i+1}+J^{\prime }\sum_{i=1}^{L-2}\left(
s_{i}^{x}s_{i+2}^{x}+s_{i}^{y}s_{i+2}^{y}\right)  \notag \\
&&+W\sum_{\alpha =x,z}\sum_{i=1}^{L}\cos \left( 2\pi ki+\varphi _{i}^{\alpha
}\right) s_{i}^{\alpha }\text{,}  \label{equ:QP}
\end{eqnarray}%
where $k=\frac{\sqrt{5}-1}{2}$(the Golden ratio), and $\varphi _{i}^{\alpha
}\in \left( 0,2\pi \right) $ is a random phase offset, the next-nearest
neighbor term is introduced to break the integrability of clean system to
stabilize the thermal phase. Without loss of generality, we choose $%
J=J^{\prime }=1$. Compared to models with random disorder, the potential in
this model is incommensurate with lattice constant while deterministic, and
hence is free of the Griffith regime\cite{Huse3}. It is now widely-accepted
the MBL transitions induced by random disorder and QP potential belong to
different universality classes, although the values of critical exponents
are under debate\cite{RD,SXZhang}.

For this QP model, we simulate three representative points in an $L=13$
system, that is, $W=1$ (thermal), $W=6$ (MBL) and $W=3.5$ (intermediate). At
each point, we compare $P_{\text{ave}}\left( r^{\left( n\right) }\right) $
to physical data, the results are shown in Fig.~\ref{fig:QP}. Perfect
fittings for thermal/MBL phase are observed as expected. While for the
intermediate region, fittings stays satisfying up to $n\leq 5$. Actually,
comparing Fig.~\ref{fig:QP} to Fig.~\ref{fig:OrthHigh}, we see $P_{\text{ave}%
}\left( r^{\left( n\right) }\right) $ works better for QP system than that
with random disorder, which can be explained as follows.

From the viewpoint of eigenvalue statistics, MBL systems with QP potential
and random disorder can be distinguished by the \emph{inter-sample randomness%
}\cite{RD,Sierant19}. To be specific, we can determine the average spacing
ratio in each sample of eigenvalue spectrum $r_{S}=\langle r\rangle _{\text{%
samp.}}$, then $P\left( r_{S}\right) $ -- the distribution of $r_{S}$ over
an ensemble of samples -- will show deviations from a Gaussian distribution
in system with random disorder, which reflects the existence of Griffiths
regime. Consequently, $V_{S}$ -- the variance of $r_{S}$ over ensemble --
will exhibit a peak at the MBL transition point, while no such peak will
appear in QP system. As we have checked, neither the mixed ensemble nor the $%
\beta $ ensemble can reproduce the peak of $V_{S}$ when varying their model
parameters $x$ and $\beta $, therefore both of them are more optimal for QP
systems, which may partially explain our observations.

\section{Conclusion and Discussion}

\label{sec5}

\bigskip We have proposed a working strategy to model the level statistics
along the thermal-MBL\ transition in random spin systems. \bigskip Our
strategy is based on two well-known random matrix (RM) models: the mixed
ensemble and Gaussian $\beta $ ensemble. We showed both models can
accurately reproduce $P\left( r^{\left( 1\right) }\right) $ with properly
chosen model parameters, while the fittings for higher-order spacing ratios
have non-negligible deviations. Specifically, the mixed ensemble
underestimates longer-range level correlations, while the opposite is true
for $\beta $ ensemble. We further show these deviations strikingly cancel
with each other by constructing their average $P_{\text{ave}}\left(
r^{\left( n\right) }\right) $, which is capable of describing level
correlation on moderate long ranges, even at the critical region. Our
results suggest the interaction range and strength between eigenvalues are
the two dominant features that responsible for the thermal-MBL transition,
in consistent with conclusions of Ref.[\onlinecite{Sierant20}].

Our strategy has both pros and cons compared to the $\beta -h$ model of Ref.[%
\onlinecite{Sierant20}]. Although our strategy works fine for fitting $%
P\left( r^{\left( n\right) }\right) $ with $n<5$ in all cases, the
deviations begin to be large for $n\geq 5$, which is outperformed by the $%
\beta -h$ model. However, one outstanding advantage of our strategy is that
the two parameters $\left( x_{\text{fit}},\beta _{\text{fit}}\right) $ in
the $P_{\text{ave}}\left( r^{\left( n\right) }\right) $ can be determined
separately, that is, $x_{\text{fit}}(\beta _{\text{fit}})$ is obtained by
fitting $P\left( r^{\left( 1\right) }\right) $ with mixed ($\beta $)
ensemble. While in $\beta -h$ model the two paramters have to be fitted
jointly, which is much more difficult to implement.

Unlike the $\beta -h$ model,\ our strategy is based on two RM models that
have well-defined parent matrix construction. It's straightforward to ask
what is the single random matrix\ model that corresponds to $P_{\text{ave}%
}\left( r^{\left( n\right) }\right) $. A natural guess would be the mixed
version of mixed ensemble and $\beta $ ensemble, that is%
\begin{equation}
M\left( x_{\text{fit}},\beta _{\text{fit}},y\right) =yM_{\alpha \rightarrow
0}\left( x_{\text{fit}}\right) +\left( 1-y\right) M_{\beta _{\text{fit}}}%
\text{.}
\end{equation}%
However, our numerical attempts find no value of $y$ will reproduce the $P_{%
\text{ave}}\left( r^{\left( n\right) }\right) $, which indicates the average
of spacing ratio distributions does not come from an linear combination of
the random matrices. The final version of a single random matrix model
remains to be explored.

Given the efficiency of $P_{\text{ave}}\left( r^{\left( n\right) }\right) $
around the critical region, it is hopeful that our strategy will be
applicable when dealing with critical phenomena of MBL transition,
especially in systems with quasi-periodic potential. There is a long debate
on whether the MBL transition induced by random disorder and quasi-periodic
potential bear identical critical exponent and hence belong to the same
universality class\cite{RD,SXZhang}. Our strategy can contribute in this
topic. A finite-size scaling study of the proper model parameters $x_{\text{%
fit}}$ and $\beta _{\text{fit}}$ around the transition region may help to
determine the critical exponent of such a transition, which may suffer less
from finite-size effect.

The physical systems studies in this work are mainly random spin chains,
while it's believed our strategy would work fine in other systems with
characteristic level statistics evolutions. For example, the disordered
Bose-Hubbard model\cite{Bose1,Bose2}, random quantum circuits\cite{Friedman}%
, interacting extended Harper model\cite{Harper1,Harper2}, and so on.

Last but not least, it is interesting to ask if the statistics of
entanglement spectrum can be modeled in the same way. Exploring this
question will help to understand the relation between the statistics of
eigenvalues to that of eigenstate wavefunction. These are all fascinating
directions for future studies.

\section*{Acknowledgements}

This work is supported by the National Natural Science Foundation of China
through Grant No.11904069.

\end{document}